\documentclass{iopart} 
\usepackage{iopams}  
\usepackage{graphicx,graphics}	
\graphicspath{{images/},{figures/}}
 \expandafter\let\csname equation*\endcsname\relax 
 \expandafter\let\csname endequation*\endcsname\relax
\expandafter\let\csname equation*\endcsname\relax 
\expandafter\let\csname endequation*\endcsname\relax
\usepackage{amsmath,amssymb,bbm} 
\usepackage{color}

\usepackage{subfig}
\usepackage{multirow}
\usepackage{array}

\def\notext#1{}

\newcommand{\identity}{\mathbbm{1}}
\newcommand{\1}{\identity}
\newcommand{\id}{\identity}

\def\1{{\mathchoice{\rm 1\mskip-4mu l}{\rm 1\mskip-4mu l}{\rm 1\mskip-4.5mu l}{\rm 1\mskip-5mu l} }}

\newcommand{\bra}[1]{{\langle #1 \vert}}

\newcommand{\ket}[1]{{\vert #1 \rangle}}

\usepackage{hyperref}

\providecommand{\openone}{\leavevmode\hbox{\small1\kern-3.8pt\normalsize1}}

\newcommand{\icf}{\address{$^{2}$Instituto de Ciencias F\'i­sicas, 
Universidad Nacional Aut\'onoma de M\'exico, Avenida Universidad s/n, 62210 Cuernavaca, Morelos, M\'exico}}
\newcommand{\pcf}{\address{$^{1}$Posgrado en Ciencias F\'i­sicas, 
Universidad Nacional Aut\'onoma de M\'exico}}
\newcommand{\uaem}{\address{$^{3}$ Centro de Investigaciones en Ciencias, Universidad Aut\'onoma del Estado de Morelos, 62209 Cuernavaca, Morelos, M\'exico}}
\newcommand{\upalermo}{\address{$^{4}$ Dipartimento di Energia, Ingegneria dell'€™Informazione e Modelli Matematici, Universit\`a  di Palermo, Viale delle Scienze, Ed. 9, 90128 Palermo, Italy}}

\begin{document} 
\title{Relations between entanglement and purity in non-Markovian dynamics} 
\author{Carlos A. Gonz\'alez-Guti\'errez$^{1,2}$, Ricardo Rom\'an-Ancheyta$^{1,2}$, Diego Espitia$^{2,3}$ and Rosario Lo Franco$^4$}
\pcf
\icf
\uaem
\upalermo

\ead{carlosgg04@gmail.com, ancheyta6@gmail.com, d.espitia@fis.unam.mx, rosario.lofranco@unipa.it}

\begin{abstract}
Knowledge of the relationships among different features of quantumness, like entanglement and state purity, is important from both fundamental and practical viewpoints. Yet, this issue remains little explored in dynamical contexts for open quantum systems.  
We address this problem by studying the dynamics of entanglement and purity for two-qubit systems using paradigmatic models of radiation-matter interaction, with a qubit being isolated from the environment (spectator configuration). We show the effects of the corresponding local quantum channels on an initial two-qubit pure entangled state in the concurrence-purity diagram and find the conditions which enable dynamical closed formulas of concurrence, used to quantify entanglement, as a function of purity. We finally discuss the usefulness of these relations in assessing entanglement and purity thresholds which allow noisy quantum teleportation. Our results provide new insights about how different properties of composite open quantum systems behave and relate each other during quantum evolutions.  
\end{abstract}


\section{Introduction}   

Dynamics of composite quantum systems interacting with their surroundings is of
central interest for understanding how quantum features are affected by the environment and for controlling them in view of their exploitation as quantum information resources~\cite{Maximilian,rivasreview,aolitareview,universalfreezing,LeggioPRA}. 
A quantum system made of two qubits is a suitable theoretical 
platform to analyze entanglement and coherence evolution from the perspective of the theory of open quantum systems \cite{petru,lofrancoreview}. 
The study of this simple system is of special relevance as it constitutes the basic building block 
for quantum gates and quantum teleportation protocols~\cite{nielsen00,benenti} which can be affected 
by external undesired interactions.

Physical models in the context of quantum optics
have been widely used in the study of {\it intrinsic} decay of quantum coherences
for one or more qubits due to the interaction with the quantized electromagnetic radiation field. 
For instance, the Jaynes-Cummings model or its generalization to a collection
of $N$ two-level atoms (or qubits) known as the Tavis-Cummings model are 
typical settings~\cite{Tavis68,finkPRL}. The study of bipartite entanglement between one qubit and the 
field~\cite{haroche2006exploring,MoyaCessaReport,leekPRL} or of multipartite entanglement among qubits \cite{amico2008RMP,Clemens:06}, 
has led to discover many interesting phenomena and also to experimental proposals for quantum protocols. 
Some examples are entanglement in simple quantum phase transitions~\cite{nielsen02},
protocols for Bell state measurements 
\cite{torres2015unambiguous}, physical implementation of quantum gates~\cite{nielsen00,dicarloNature,PhysRevLett.108.120501} and generation of quantum correlations in qubit networks~\cite{Solano1}.

One of the main drawbacks encountered when two qubits locally interact with their own Markovian (memoryless) environment is the so-called
entanglement sudden death (ESD), that is the complete disappearance of entanglement at a finite time~\cite{Yu04,Cui07,Yonac1,Yonac2,EberlyReview,PinedaPRL2007,PinedaNew}. 
This phenomenon, whose experimental evidence has been also proved~\cite{Almeida579,kimble2007PRL}, has then motivated the development of efficient strategies to avoid it~\cite{Lim14,lofrancoQIP,lofrancoPRB,adeline2014,darrigo2012AOP,bellomo2008trapping,bellomo2009ASL,bellomo2007PRL} or delay it~\cite{gonzalez2015,manSciRep,LoFrancoNatCom,bellomo2010PhysScrManiscalco,bellomoSavasta2011PhyScr,bellomo2012noisylaser,manPRA}, typically based on suitable non-Markovian (memory-keeping) environments and local operations.

The scope of this paper is to provide new insights about the non-Markovian dynamics of the quantum correlations captured by entanglement from the perspective of its interplay with purity, which identifies the degree of mixedness of a quantum state being related to coherence. Such a study is still little addressed \cite{Ziman-CP05,PhysRevA.72.042320}, particularly in the presence of local non-Markovian environments \cite{mazzolapalermo2010PRA}. Knowledge of relationships between entanglement and purity in specific dynamical contexts is important not only from a fundamental viewpoint but also from a practical one. In fact, it would provide quantitative thresholds of entanglement for a given purity at a certain time which allow quantum protocols, like teleportation \cite{qinQIP,roarXiv}, entanglement swapping \cite{roaPRA} and entanglement percolation \cite{acin2007NatPhys}. 
To this aim our strategy is to consider a two-qubit {\em central system} using the so-called {\it spectator configuration} \cite{pinedaTesis}, where one of the qubits is isolated and acts as a {\it probe}. 
This idealized configuration is a convenient way to
investigate non-trivial dynamics of entanglement versus purity for two qubits without
any type of interaction between them. 
On the other hand, the characterization of quantum processes under particular channels 
or operations within this simple open quantum system can be implemented experimentally. 
Realizations of unital and nonunital (both Markovian and non-Markovian) channels acting on one of two qubits 
are posible using all-optical setups \cite{adeline2014,shaham2015entanglement,LoFrancoNatCom,chiuri2012} and are also achievable in circuit QED devices \cite{finkPRL,leekPRL}. 
For our analysis we shall also employ the concurrence-purity ($C$-$P$) plane~\cite{Ziman-CP05}, which is a 
powerful tool that brings a general overview of the system dynamics and it is not commonly used in quantum optics literature.
We focus on three models which shall allow us to obtain exact analytical results with a consequent better understanding of the system evolution, namely: Tavis-Cummings (TC), Buck-Sukumar (BS) and spin-boson (SB) models. 

The paper is structured as follows. In 
Sections \ref{two::Tavis::Cummings}, \ref{TABSM} and \ref{TASBM} we
analyze the two-qubit TC, BS and SB models, respectively, for which we obtain the
exact time evolution for the reduced density operator of the central system.
Exact expressions for purity and concurrence are also derived. 
In Sec.~\ref{R&D} we discuss the results and explore the $C$-$P$ diagram identifying the 
nature of different decoherence processes in the two-qubit central system and their use for implementing noisy teleportation. 
Finally, in Sec.~\ref{Conclusions} we give our conclusions.

\section{Two-qubit Tavis-Cummings model}\label{two::Tavis::Cummings}
The interaction between two identical two-level atoms (qubits) $A$ and $B$ with a single
mode of the electromagnetic radiation field with frequency $\nu$ in the dipole and rotating-wave
approximations is described by the following Tavis-Cummings Hamiltonian \cite{Tavis68}
(we set $\hbar=1$)
\begin{equation}
\label{two:atoms:tavis:spectator}
  H_{\texttt{TC}}=\frac{\omega_0}{2}(\sigma_{z}^{A}+\sigma_{z}^{B})+\nu a^{\dagger}a+
  \sum_{j=A,B}g_{j}\big(a\sigma_{+}^{j}+a^{\dagger}\sigma_{-}^{j}\big),
\end{equation}
where $a$, $a^{\dagger}$  are the usual bosonic operators satisfying $[a,a^{\dagger}]=1$,
$\sigma_z^j$ is the $z$-component of Pauli matrices and
$\sigma_{\pm}^{j}$ are the rising and lowering operators for atoms $A$ and $B$. 
We remark that this model is experimentally realized in circuit QED \cite{finkPRL}.
Here we will focus on the particular case where $g_{B}=0$ and $g_{A}=g$, i.e., 
only one of the atoms is interacting with the field. This setting could be
realized with the atom $B$ outside of the cavity \cite{ZhiJian} or with it in a 
node of the electromagnetic field. 
In this context the atom $B$ acts as a {\it probe} from which one can 
obtain information about the other systems (atom $A$ and/or the field).
This is the spectator configuration.
For simplicity, in the following we restrict our analysis to the resonant case
$\omega_{0}=\nu$. 
With the aforementioned considerations it is easy to obtain the exact time
evolution operator for the Hamiltonian of Eq.~\eqref{two:atoms:tavis:spectator}
in the interaction picture and in the atomic basis
$\{\ket{ee},\ket{eg},\ket{ge},\ket{gg}\}$, which reads
\begin{align}
U(t)_{\texttt{TC}}=
\begin{pmatrix}
\cos(gt\sqrt{aa^\dagger})&
-\rmi V \sin(gt\sqrt{a^\dagger a})\\
-\rmi V^\dagger\sin(gt\sqrt{aa^\dagger})&
\cos(gt\sqrt{a^\dagger a})
\end{pmatrix}\otimes\id_B,
\end{align}
where $\id_B$ is the identity operator for the qubit $B$ Hilbert space, and
we have used the well known expression for the Jaynes-Cummings (JC) time propagator
\cite{MoyaCessaReport} in terms of the Susskind-Glogower operators defined as \cite{Susskind} 
\begin{align}\label{london-operators}
V=\frac{1}{\sqrt{a^\dagger a+1}}a=\sum_{n=0}^\infty|n\rangle\langle n+1|,\,\
V^\dagger =a^\dagger\frac{1}{\sqrt{a^\dagger a+1}}=\sum_{n=0}^\infty|n+1\rangle\langle n|.
\end{align} 
These operators are non-unitary and satisfy the commutation relation
 $[V,V^\dagger]=|0\rangle\langle 0|$.

In order to investigate the reduced dynamics of the two-qubit system we assume
the total initial state as a product state
$\rho(0)=\varrho_{\Psi}(0)\otimes\rho_{f}(0)$ where
\begin{align}
\label{initial:state}
\varrho_{\Psi}(0)=\frac{1-x}{4}\identity+x\ket{\psi}\bra{\psi},
\end{align}
is a Werner-like state for the central system with purity parameter $x\in[0,1]$,  
$\ket{\psi}=\sin{\phi}\ket{ee}+\cos{\phi}\ket{gg}$ and
$\rho_{f}(0)$ is an arbitrary initial state of the field. Such a state reduces to a Bell-like state when $x=1$ and is contained in a wider class of  two-qubit states known as X states, which are represented by a density matrix having only diagonal and off-diagonal terms different from zero \cite{lofrancoreview}.
We focus on two particular field states of interest: the number state and coherent state, which represent the most quantum and the most classical states of the radiation field, respectively. 

\subsection{Field in a number state}
In this case we consider the field to be initially in a pure state with a
definite number of photons, i.e.,  $\rho_{f}(0)=\ket{n}\bra{n}$.
Using this field state in $\rho(0)$
and tracing over the degrees of freedom of the field we can get the reduced
density operator for the central system as
\begin{equation}\label{reduce_densiy_matrix}
\varrho(t)=\tr_f[U(t)\rho(0)U^{\dagger}(t)].
\end{equation}
For simplicity we write down only the non zero matrix elements for the reduced density operator
\begin{align}\label{matrix:elementsJCM}
\varrho_{11}(t)&= \Big(\frac{1-x}{4}+x\sin^{2}\phi\Big)\cos^{2}(gt\sqrt{n+1})+\frac{1-x}{4}\sin^{2}({gt\sqrt{n}}), \nonumber \\
\varrho_{22}(t)&=\Big(\frac{1-x}{4}+x\cos^{2}\phi\Big)\sin^{2}({gt\sqrt{n}})+\frac{1-x}{4}\cos^{2}(gt\sqrt{n+1}), \nonumber \\
\varrho_{33}(t)&=\Big(\frac{1-x}{4}+x\sin^{2}\phi\Big)\sin^{2}(gt\sqrt{n+1})+\frac{1-x}{4}\cos^{2}(gt\sqrt{n}), \\
\varrho_{44}(t)&=\Big(\frac{1-x}{4}+x\cos^{2}\phi\Big)\cos^{2}({gt\sqrt{n}})+\frac{1-x}{4}\sin^{2}(gt\sqrt{n+1}), \nonumber \\
\varrho_{14}(t)&=x\sin{\phi}\cos{\phi}\cos({gt\sqrt{n+1}})\cos({gt\sqrt{n}}),\,\,\,
\varrho_{41}(t)={\varrho_{14}(t)}^{*}.\nonumber
\end{align}
Notice that the reduced density operator maintains during the time evolution
its initial X structure. With the reduced density matrix of Eq.~\eqref{matrix:elementsJCM}
we can calculate at any time the evolution of purity and concurrence for the central 
system which are standard measurements of decoherence and entanglement. 

To quantify the loss of coherence trough the degree of mixedness of the two-qubit system we use the purity of a density
operator which is defined as
\begin{equation}\label{purity}
 P(t)=\tr[\varrho(t)^2].
\end{equation}
The purity takes its maximum value of one if the state is a one-dimensional projector, i.e. if it is a pure state. 
The minimum value of this quantity is bounded by the inverse of the dimension of the system Hilbert space.

The entanglement shared between two qubits can be quantified using the concurrence, which is defined
for a general mixed state $\varrho$ as~\cite{Wootters98}
\begin{equation}\label{concurrencia}
C(\varrho)=\max\{0,\tilde\lambda_1-\tilde\lambda_2-\tilde\lambda_3-\tilde\lambda_4\},
\end{equation}
where $\tilde\lambda_{i}$ are the square roots of the eigenvalues of $\varrho\tilde\varrho$ 
in non-increasing order. The operator $\tilde\varrho$ is obtained
by applying a spin flip operation on $\varrho$, i.e,
$\tilde\varrho=(\sigma_{y}\otimes\sigma_{y})\varrho^{*}(\sigma_{y}\otimes\sigma_{y})$ and the complex conjugate 
is taken in the atomic basis of the two qubits.\medskip\\
For a X state of the form of Eq.~(\ref{matrix:elementsJCM}), the concurrence can be easily obtained via
 \cite{Ziman-CP05}
 \begin{equation}\label{concurrencia_X}
  C(\varrho_{\text{X}})= 2 \max\{0,|\varrho_{14}|-\sqrt{\varrho_{22}\varrho_{33}}\}.
 \end{equation}
To get easy to handle explicit expressions of the quantifiers, we analyze the particular case with $x=1$ and $\phi=\pi/4$, which 
corresponds to an initial pure Bell state of the two-qubit system. 
A straightforward calculation shows that purity and concurrence read
\begin{equation}
\label{purity:JCM:spectator}
P(t)= \frac{1}{2}+\frac{1}{8}\left[4\cos^{2}({gt\sqrt{n}})\cos^{2}({gt\sqrt{n+1}})-1\right]+\frac{1}{16}\left[\cos{(4gt\sqrt{n})}+\cos{(4gt\sqrt{n+1}})\right].
\end{equation}
\begin{equation}
\label{concu:spectator:JCM}
 C(t)=2\max\{0,\frac{1}{2}\left(\left|\cos({gt\sqrt{n}})\cos({gt\sqrt{n+1}})\right|-
 \left|\sin({gt\sqrt{n}})\sin({gt\sqrt{n+1}})\right|\right)\}.
\end{equation}
We notice that for $n=0$ (vacuum field state), purity and concurrence are related via
\begin{equation}
\label{c-p::homogenization}
C(t)=\sqrt[4]{2P(t)-1},
\end{equation}
which is the typical behaviour that characterizes a homogenization process in a $C$-$P$ diagram \cite{Ziman-CP05} and tells us that the two qubits are entangled whenever the purity is larger than $1/2$.
This process belongs to a class of non-unital channels (see Sec.~\ref{R&D} for details).

\begin{figure}
\centering
\begin{tabular}{cccc}
\hspace*{-1.0cm}\subfloat[]{\includegraphics[width=.55\linewidth]{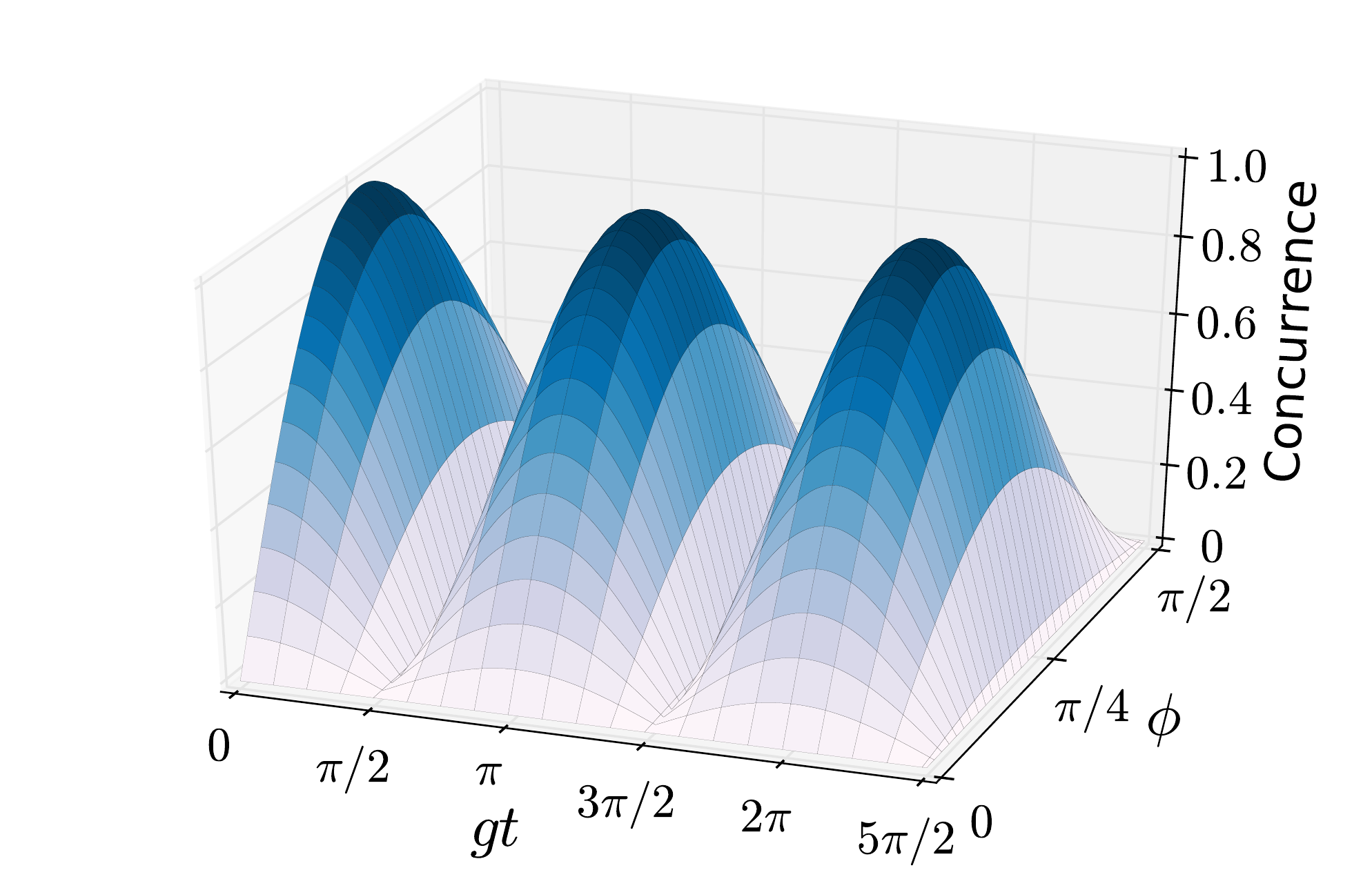}} & 
\hspace*{-0.7cm}\subfloat[]{\includegraphics[width=.55\linewidth]{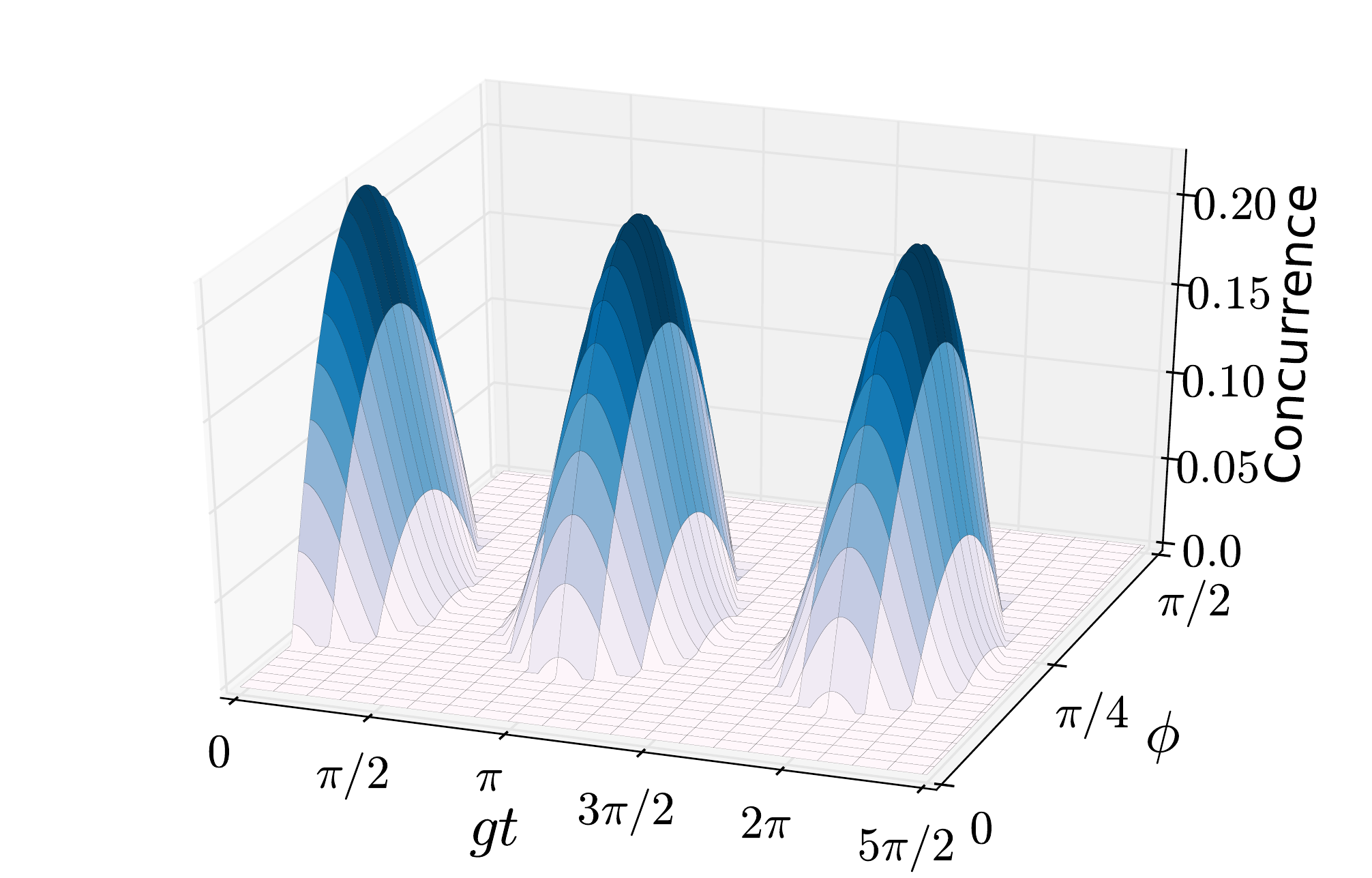}} & \\
\end{tabular}
\caption{Concurrence as a function of scaled time $gt$ and  
initial degree of entanglement $\phi$ for the spectator two-qubit TC model 
in the vacuum state $n=0$. Two cases are shown:  (a) with purity parameter $x=1$
there is vanishing of entanglement at $gt=(m+1/2)\pi$, (b) For $x=0.48$ collapses 
and revivals of entanglement are observed. This behavior shows a strong dependence on the initial conditions, 
as reported in Ref.~\cite{Cui07}.}\label{concu:x:uno}
\end{figure}

In Fig.~\ref{concu:x:uno} we show the evolution of concurrence by
substituting the matrix elements of Eq.~\eqref{matrix:elementsJCM} in Eq.~\eqref{concurrencia_X}
with the field in the vacuum state $n$=$0$ and an arbitrary initial degree of entanglement.
The figure shows two cases: (a) pure state ($x=1$), for which $C(t)=2\max\{0,|\cos\phi\sin\phi\cos gt|\}$ and (b) mixed state ($x=0.48$).
The time behaviors are in accordance with the non-dissipative case of a single qubit subject 
to a single-mode radiation field in the vacuum state (zero temperature perfect cavity). 

\subsection{Field in a coherent state} \label{sub::coherent_state}
\begin{figure}[t!]
 \centering
 \includegraphics[width=8cm, height=5cm]{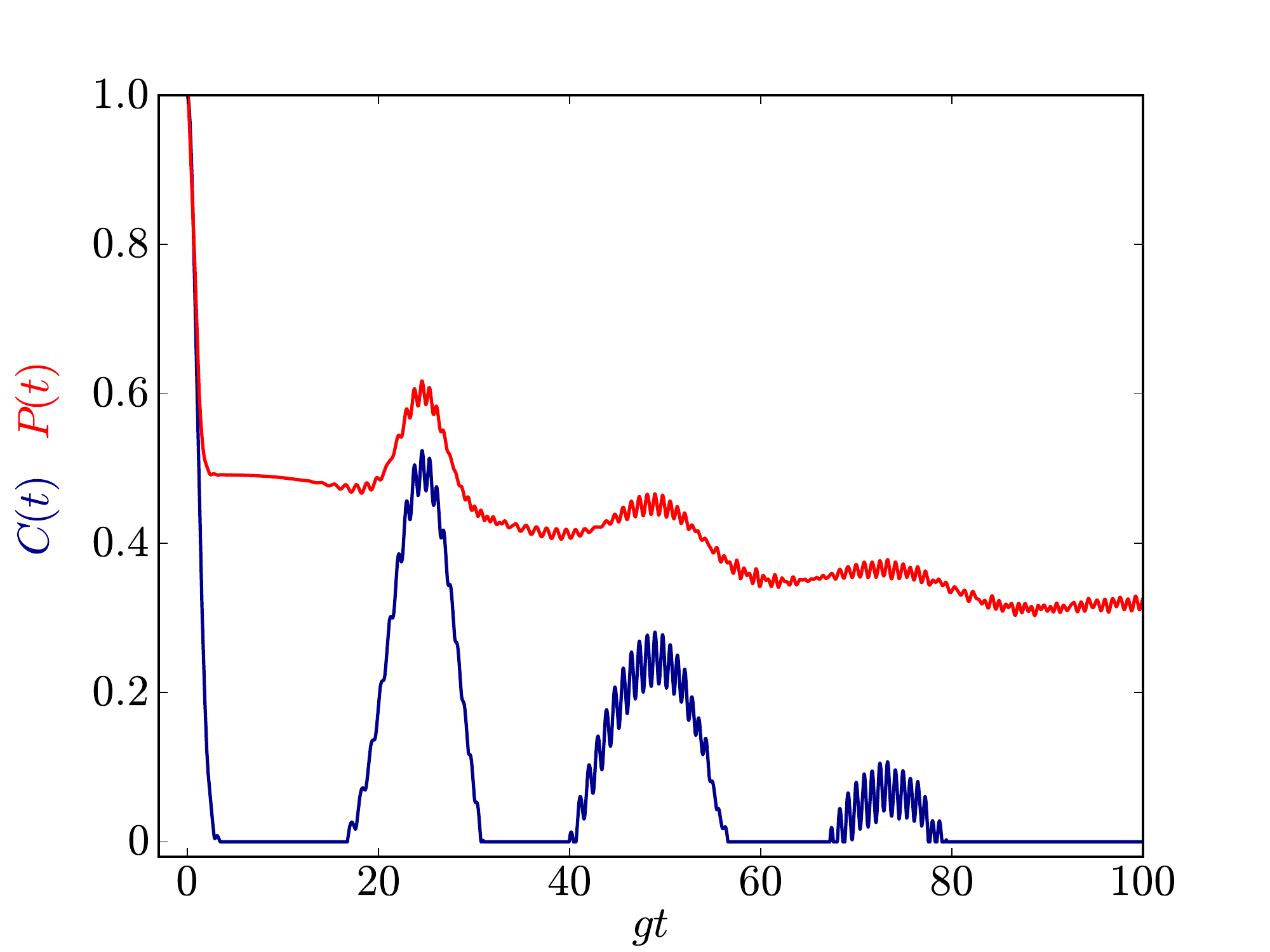}
 \caption{Purity (upper red line) and concurrence (lower dark blue line) as functions of scaled time $gt$
 for the spectator two-qubit TC model. Bell ($x=1$, $\phi=\pi/4$) and coherent ($\bar{n}=15$) initial states
  were used. Concurrence shows collapses and revivals of entanglement with the envelope eventually decaying
 at $gt_r\approx2\pi\sqrt{\bar n}$.}
 \label{fig:Pu_Con_JCM_Coh}
\end{figure}

We now choose the initial radiation field
in a coherent state, which is a typical situation in cavity-QED experiments \cite{haroche2006exploring}.
In this case the field state is given by $\rho_{f}(0)$=$\ket{\alpha}\bra{\alpha}$,
where $\ket{\alpha}$=$\sum_{m=0}^\infty C_{m}\ket{m}$ 
with $C_{m}$=$\exp(-|\alpha|^2/{2}){\alpha^{m}}/{\sqrt{m!}}$.
The explicit elements of the reduced density operator for
 $x=1$ and $\phi=\pi/4$ are
\begin{align}\label{matrix:elementsJCM:Coherent}
\varrho_{11}&= \sum_{m}|C_{m}|^2\cos^{2}({gt\sqrt{m+1}}), 
\varrho_{13}=\text{i}\sum_{m}C^{*}_{m+1}C_{m}\sin({gt\sqrt{m+1}})\cos({gt\sqrt{m+2}}),\nonumber \\
\varrho_{12}&= \frac{\text{i}}{2}\sum_{m}C^{*}_{m+1}C_{m}\sin({2gt\sqrt{m+1}}), 
\,\,\,\,\,\varrho_{14}=\sum_{m}|C_{m}|^2\cos({gt\sqrt{m}})\cos({gt\sqrt{m+1}}), \nonumber \\
\varrho_{22}&= \sum_{m}|C_{m}|^2\sin^{2}({gt\sqrt{m+1}}), 
\,\,\varrho_{23}=\sum_{m}C^{*}_{m}C_{m+2}\sin({gt\sqrt{m+1}})\sin({gt\sqrt{m+2}}), \nonumber \\
\varrho_{33}&=\sum_{m}|C_{m}|^2\sin^{2}({gt\sqrt{m}}), 
\,\,\,\,\,\,\,\,\,\,\, \varrho_{24}=-\text{i}\sum_{m}C^{*}_{m}C_{m+1}\sin({gt\sqrt{m+1}})\cos({gt\sqrt{m}}), \nonumber \\
\varrho_{44}&= \sum_{m}|C_{m}|^2\cos^{2}({gt\sqrt{m}}), 
\,\,\,\,\,\,\,\,\,\,\, \varrho_{34}=-\frac{\text{i}}{2}\sum_{m}C^{*}_{m+1}C_{m}\sin({2gt\sqrt{m+1}}),
\end{align}
where we have omitted the explicit time dependence in the matrix elements $\varrho_{jk}(t)$.
As in the standard JC model, the sums in Eq.~\eqref{matrix:elementsJCM:Coherent}
cannot be evaluated in a closed form, so analytical expressions for purity and 
concurrence are too cumbersome to be shown here. 
In Fig.~\ref{fig:Pu_Con_JCM_Coh} we then show plots of purity and concurrence as functions of time, where
the field state is initially in a coherent state with average photon number 
$\bar{n}=|\alpha|^2=15$. Differently from the previous case of initial number state, now entanglement and purity eventually decay presenting oscillations during the evolutions. We point out that purity peaks follow entanglement revivals which however does not mean that the larger the purity (or smaller the mixedness), the larger the entanglement. This can be immediately seen by comparing, for instance, the behaviors at the time regions $2<gt<18$ (zero entanglement) and $68<gt<78$ (entanglement revival).

\section{Two-qubit Buck-Sukumar model}\label{TABSM}
In this section we consider a variant of the model studied in Sec.~\ref{two::Tavis::Cummings} which is inspired to the so-called Buck-Sukumar (BS) model~\cite{Buck-Sukumar}.
In that work the authors propose an exactly solvable qubit-field Hamiltonian which 
is useful to describe nonlinear interactions.
The Hamiltonian for the two-qubit BS model in the spectator configuration is given by 
\begin{equation}\label{two:atoms:BS:spectator}
 H_{\texttt{BS}}=\frac{\omega_{0}}{2}\left(\sigma^{A}_{z}+\sigma^{B}_{z}\right)+\nu a^{\dagger}a+
 g\left(a\sqrt{N}\sigma^{A}_{+}+\sqrt{N}a^{\dagger}\sigma^{A}_{-}\right),
\end{equation}
where $N=a^\dagger a$. Unlike Eq.~(\ref{two:atoms:tavis:spectator}) 
this model allows an intensity-field dependent coupling.
In the resonant case the time evolution operator in the interaction picture is
\begin{eqnarray}
U_{\texttt{BS}}(t)=
\begin{pmatrix}
\cos\left[gt(N+1)\right]&-\rmi V\sin\left[gtN)\right]\\
-\rmi V^\dagger\sin\left[gt(N+1)\right]&\cos\left[gtN\right]
\end{pmatrix}\otimes\id_B.
\end{eqnarray}
Using the same initial condition for the two-qubit system Eq.~\eqref{initial:state}, the
matrix elements for the reduced density operator are analogous to Eqs.~(\ref{matrix:elementsJCM}) and (\ref{matrix:elementsJCM:Coherent})
(except for the square root in the trigonometric functions argument, i.e. $\sqrt{x}\to x$)
for the field in a number and coherent state respectively.

\begin{figure}
\begin{center}
{\subfloat[]{\includegraphics[width=8cm, height=5cm]{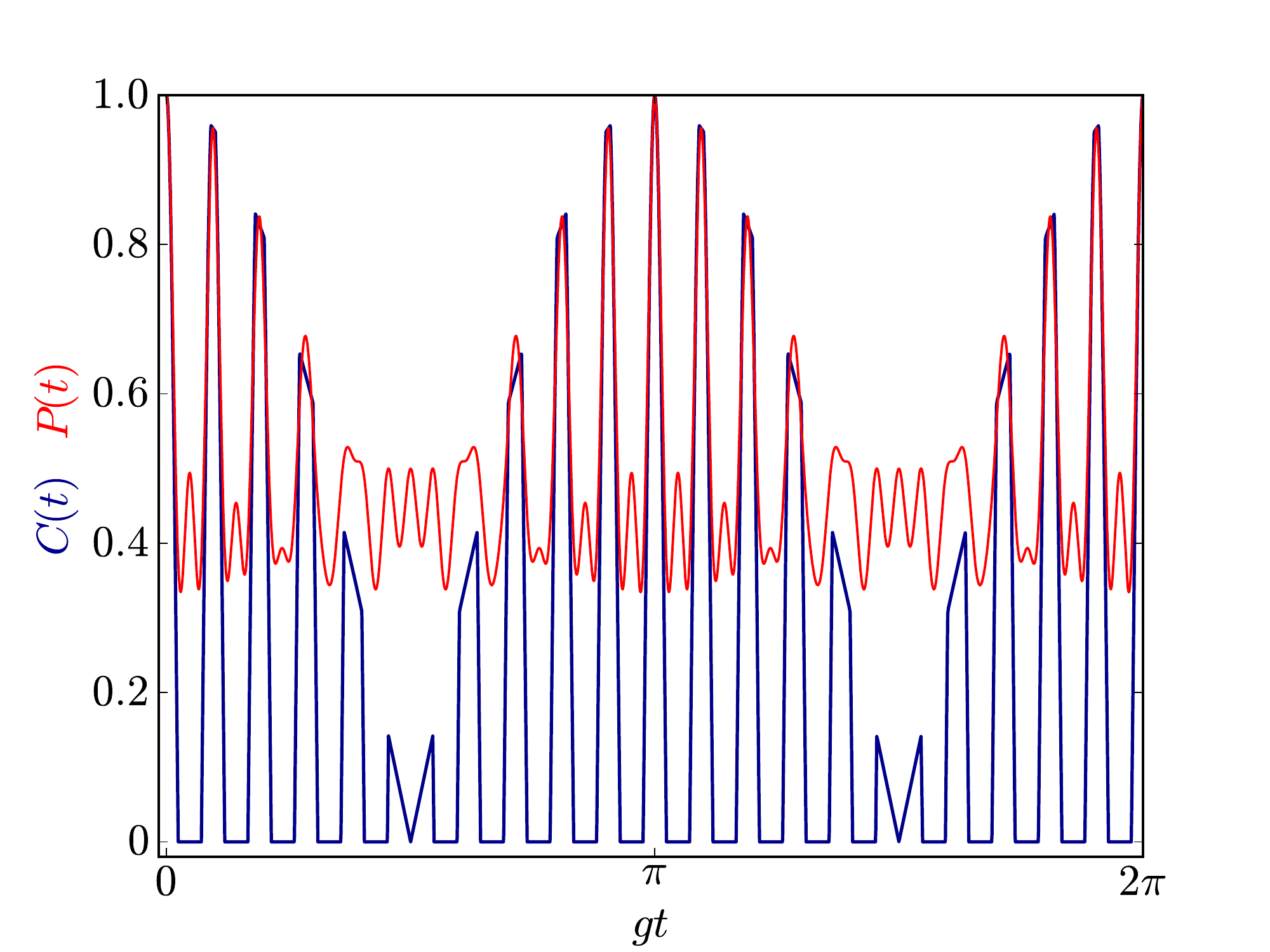}} \\
\subfloat[]{\includegraphics[width=8cm, height=5cm]{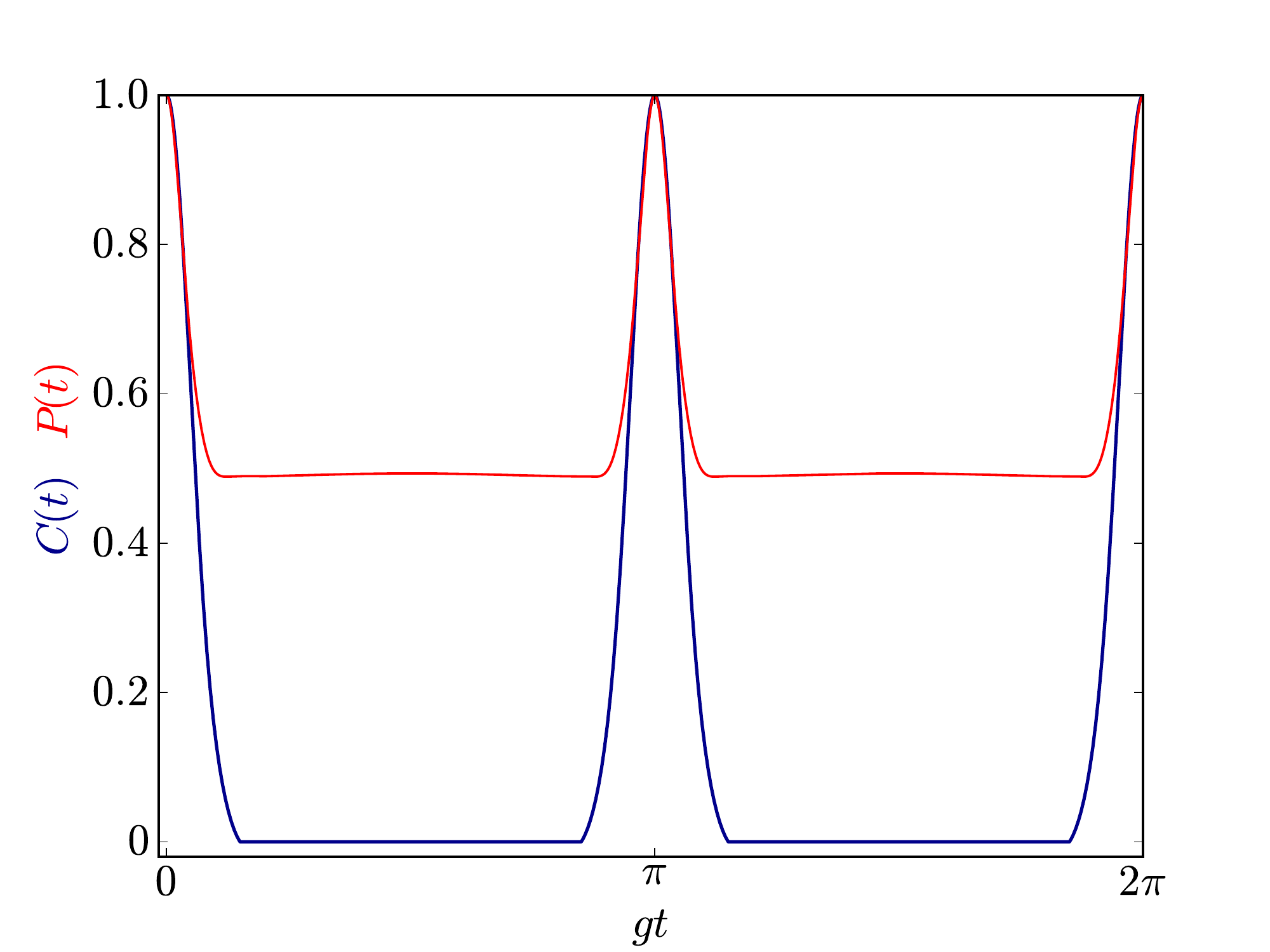}}}
\end{center}
\caption{Purity (upper red line) and concurrence (lower dark blue line) evolution for the 
spectator two-qubit BS model. The two qubits start in a Bell state. The field starts in: 
(a) number state with $n=10$ and (b) coherent state with $\bar{n}=10$. 
A sequence of entanglement dark periods and complete entanglement recoveries occur in both cases due to the nonlinear
interaction.}
\label{fig:con:BS:10Fock}
\end{figure}

Purity and concurrence for the Bell pair ($x$=$1$, $\phi$=$\pi/4$) with the field 
starting in the number state $|n\rangle$ are
\begin{equation}\label{purtiy:spectator:BSM}
P(t)= \frac{1}{2}+\frac{1}{8}\left(4\cos^{2}\left[{gt{n}}\right]\cos^{2}\left[{gt(n+1)}\right]-1\right)+\frac{1}{16}\left(\cos\left[4gt{n}\right]+\cos\left[4gt(n+1)\right]\right),
\end{equation}
\begin{equation}\label{concu:spectator:BSM}
C(t)=2\max\{0,\frac{1}{2}\left(\left|\cos\left[{gt{n}}\right]\cos\left[{gt(n+1)}\right]\right|-
\left|\sin\left[{gt{n}}\right]\sin\left[{gt(n+1)}\right]\right|\right)\}.
\end{equation}
In Fig.~\ref{fig:con:BS:10Fock}(a) we have plotted Eqs.~(\ref{purtiy:spectator:BSM}) 
and (\ref{concu:spectator:BSM}) as functions of time with $n=10$ photons.
A behaviour similar to that of Fig.~\ref{fig:con:BS:10Fock}(a) is found between an isolated atom
and a Jaynes-€"Cummings atom \cite{ZhiJian}.

On the other hand, when the field is initially in a coherent state analytical
expressions for $P(t)$ and $C(t)$ in the two-qubit BS model are cumbersome, as pointed out in the previous section, and we limit to report their plots. Evolutions of purity and
concurrence for this case are displayed in Fig.~\ref{fig:con:BS:10Fock}(b) as functions of scaled time 
for $x=1$, $\phi=\pi/4$ and $\bar n=10$. We highlight that now, in contrast to what happened in the JC model with an initial coherent field state (see Fig.~\ref{fig:Pu_Con_JCM_Coh}), a complete spontaneous recovery of the initial entanglement can be found due to the nonlinear atom-field interaction. Purity and entanglement again show the same qualitative behavior but now larger values of purities always correspond to larger values of entanglement ($P=1/2$ when $C=0$ in the plateaux and $P=1$ when $C=1$ in the peak).  

\section{Two-qubit spin-boson model}\label{TASBM}
\begin{figure}[t!]
 \centering
 \includegraphics[width=8cm, height=5.5cm]{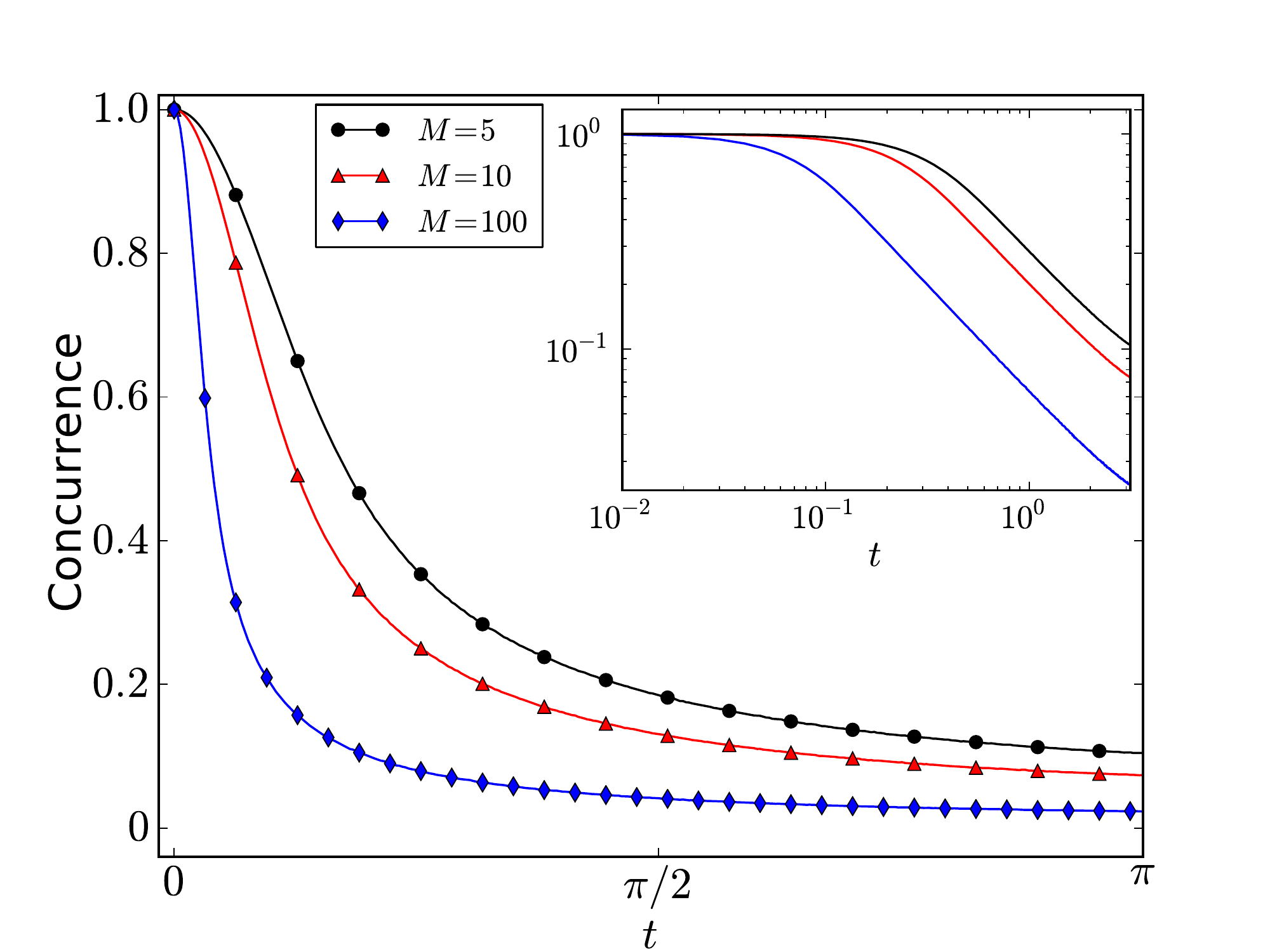}
 \caption{Entanglement evolution for the spectator two-qubit SB model.
 An ensemble average over $10^6$ samples was realized using Eq.~(\ref{concuSBM}), $x$=$1$,
 $\phi$=$\pi/4$, $g_j$ and $\omega_j$ were chosen randomly from the interval $[0,1]$, with
 $M$= 5 (black), 10 (red), 100 (blue). The inset shows in log scale the Gaussian 
 (exponential) entanglement decay for short (long) times.}
 \label{fig:Con_SB}
\end{figure}

The two-qubit spin-boson model describes two spin $1/2$ particles coupled to 
an environment of $M$ non-interacting quantum harmonic oscillators~\cite{Maximilian}, which can be experimentally realized in cavity and circuit QED \cite{petru,finkPRL} and also simulated by all-optical setups with Sagnac interferometers \cite{Almeida579,chiuri2012}.
The pure-dephasing Hamiltonian in the spectator scheme is given by
\begin{equation}\label{spinboson}
H_{\texttt{SB}} = \frac{\omega_0}{2} \left(\sigma_z^{A} + \sigma_z^{B}\right) +
\sum_{j=1}^M \omega_j a_j^\dagger a_j + 
\sigma_z^{A} \otimes \sum_{j=1}^M\big(g_j a_j^\dagger + g^*_j a_j\big).
\end{equation}
Notice that the qubit-environment linear coupling term is an energy conserving interaction
since the central system Hamiltonian commutes with $H_{\texttt{SB}}$. The corresponding time evolution operator
in the interaction picture is
\begin{align}
U_{\texttt{SB}}(t)=
\begin{pmatrix}
\prod_j D(\lambda_j(t))&
0\\
0&
\prod_j D(-\lambda_j(t))
\end{pmatrix}\otimes\id_B,
\end{align}
where ${D}\left({\lambda_j(t)}\right) \equiv \exp [(\lambda_j(t)\hat{a}_j^\dagger - \lambda_j^*(t)\hat{a}_j)]$ 
is the usual Glauber displacement operator for each mode and
 $\lambda_j(t) \equiv(g_j/\omega_j)[1-\exp(\mathrm{i}\omega_j t)]$.
If we set all the oscillators in the ground state  $\rho_f(0)$=$\bigotimes_{j=1}^M|0\rangle_j\,_j\langle 0|$
and the two-qubit system in $\varrho_\Psi(0)$, the non-zero matrix elements of 
the total density operator in the atomic basis are
\begin{align}\label{matrix:elementsSBM}
\rho_{11}(t)&=\Big(\frac{1-x}{4}+x\sin^{2}\phi\Big)\prod_j |{\lambda_j(t)}\rangle \langle\lambda_j(t)|, 
\quad\,\,\,\rho_{33}(t)=\frac{1-x}{4}\prod_j|-{\lambda_j(t)}\rangle \langle{-\lambda_j(t)}|, \nonumber \\
\rho_{44}(t)&=\Big(\frac{1-x}{4}+x\cos^{2}\phi\Big)\prod_j|-{\lambda_j(t)}\rangle \langle-{\lambda_j(t)}|, 
\,\,\quad\rho_{22}(t)=\frac{1-x}{4}\prod_j |{\lambda_j(t)}\rangle\langle{\lambda_j(t)}|, \nonumber \\
\rho_{14}(t)&=\rho_{14}(t)^*=x\sin\phi\cos\phi \prod_j |{\lambda_j(t)}\rangle \langle{-\lambda_j(t)}|,
\end{align}
where $|\lambda_j(t)\rangle$$\equiv$$D(\lambda_j(t))|0\rangle_j$
is the coherent state for the $j$-th oscillator. 
As we have done in previous sections, we trace out over 
the environment in order to obtain the reduced density operator of the central system:
$\varrho_{11}(t)$=$($1$-$$x$$)/4$+$x\sin^2\phi$, $\varrho_{22}(t)$=$\varrho_{33}(t)$=$(1$$-$$x)/4$,
$\varrho_{44}(t)$=$(1$$-$$x)/4$+$x\cos^2\phi$ and
$\varrho_{14}(t)$=$\varrho_{41}(t)$=$x\sin\phi\cos\phi\exp [-\Gamma(t)]$,
where $\Gamma(t)$=$\sum_j{4|g_j|^2}(1$$-\cos(\omega_i t))/\omega_j^2$ is the decoherence factor.
From the Eqs.~\eqref{purity} and \eqref{concurrencia_X} it is trivial to 
obtain purity and concurrence for the central system. For instance, the explicit expression for concurrence is 
\begin{equation}\label{concuSBM}
C(t)=\max\big\{0,x|\sin2\phi|\mathrm{e}^{-\Gamma(t)}-{(1-x)}/{2}\big{\}}.
\end{equation}
For $x$=$1$ purity and concurrence are related via
\begin{equation}\label{C-P_SBM}
C(t)=\sqrt{2P(t)-2(\sin^4\phi+\cos^4\phi)},
\end{equation}
which is a generalized form of the expression describing a dephasing process
induced by a local operation acting on a Bell state given by $C=\sqrt{2P-1}$.
From Eq.~(\ref{concuSBM}), one finds that entanglement vanishes whenever $\Gamma(t)=-\ln((1-x)/(2x|\sin2\phi|))$.  
Assuming all the modes to be identical $(g_j$=$g$, $\omega_j$=$\omega)$,
with $\phi$=$\pi/4$, the time when entanglement disappears is 
$t_{d}=\arccos\big[1+\frac{1}{M}\frac{\omega^2}{4|g|^2}\ln((1-x)/(2x))\big]$. 
In Fig.~\ref{fig:Con_SB} we plot concurrence of Eq.~\eqref{concuSBM} as a function 
of time for several realizations of $g_j$ and $\omega_j$ which are randomly chosen 
from interval $[0,1]$. $M$ stands for different dimensions of the environment. We emphasize that this time behavior is non-Markovian meaning that its decay is not exponential at short times, a situation reminiscent of pure-dephasing evolution in the solid state due to inhomogeneous broadening \cite{PhysRevLett.94.167002,lofranco2012PhysScripta}. 

\section{Discussions}\label{R&D}

In this section we discuss the results for the evolution of purity and concurrence 
for the different models studied in previous sections.

\subsection{General aspects on the time behaviors}
For the two-qubit TC model with the field starting in the vacuum state, concurrence (and also purity) 
is a periodic function of time as can be seen in Fig.~\ref{concu:x:uno}. We have explored two different
initial conditions for the two-qubit system~\eqref{initial:state}: pure entangled state 
Fig.~\ref{concu:x:uno} (a) and entangled state with a degree of mixedness Fig.~\ref{concu:x:uno}(b).
In both cases we observe the expected decay of correlations at short times in the initial 
entangled state due to the interaction with the field. Fig.~\ref{concu:x:uno}(a) shows 
complete entanglement revivals at times given by $gt=n\pi$. A similar behaviour is shown 
in Fig.~\ref{concu:x:uno}(b) but in this case the entanglement remains zero for finite 
intervals of time, identified as entanglement dark periods \cite{lofrancoreview}, 
followed by complete entanglement recoveries as time goes by.
In the case under consideration, the TC interaction permits only zero photons or
one photon to reside in the cavity, i.e., the cavity acts effectively as a two-level system,
so the Hilbert space available for the environment is finite and gives rise to entanglement 
rebirths in the central system. When entanglement completely disappears in the central system, quantum correlations
must be contained in other bi-partitions \cite{lopez2008PRL,lopez2010PRA}, for instance between the isolated qubit and 
the field or the central system and the field. This effective three-qubit system is a convenient
framework for understanding the dynamical mechanisms of entanglement sharing among the parts of a composite system with a quantum reservoir \cite{lofrancoreview,darrigo2012AOP}.
Dynamical behaviors qualitatively similar to those obtained in the case when both qubits are open \cite{Cui07,lofrancoreview} have been here found. This implies that the spectator configuration is able to reproduce general dynamical features exhibited by more complex systems, provided that each qubit of the system is locally interacting with its own environment. 

Concerning the second initial condition for the environmental state in the TC model, i.e. the field
prepared in a coherent state, we notice that this is the situation in which the Hilbert space
is formed by an infinite basis of number states. In principle it is possible 
that entanglement can be shared in arbitrary multipartitions of the Hilbert space not allowing 
the complete backflow of information to the central system. 
This sort of local coherent-state control leads to revivals of entanglement whose amplitude eventually decays, 
as predicted for the case of two open qubits~\cite{Yonac1,Yonac2}.       
Purity and concurrence evolution for the central system have been plotted in Fig.~\ref{fig:Pu_Con_JCM_Coh}
when the average photon number of the field coherent state is $\bar{n}=15$. 
Both quantities oscillate but the periodicity in both quantities is no longer maintained.
This time behaviour resembles the evolution of the atomic inversion in the standard one-qubit JC model 
where non-complete revivals are consequence of constructive quantum interference between 
states in the Fock basis~\cite{haroche2006exploring}. Since we have used the spectator configuration
it is easy to see that the time of entanglement revival is given by $gt_{r}\approx2\pi\sqrt{\bar n}$. 

Successively, considering the intensity-dependent field interaction described by the two-qubit BS Hamiltonian 
in Eq.~(\ref{two:atoms:BS:spectator}), we have plotted purity and concurrence with the field in
a number ($n=10$) and a coherent state ($\bar{n}=15$) in Figs.~\ref{fig:con:BS:10Fock}(a)
and \ref{fig:con:BS:10Fock}(b), respectively. 
In contrast to what was observed for the TC model, $C(t)$ and $P(t)$ are now 
$\pi$-periodic functions independent of the number of photons. 
Interestingly, when the radiation field is initially in a coherent state there are complete entanglement revivals
(see Fig.~\ref{fig:con:BS:10Fock}(b)) regardless that we are dealing with an infinite number of 
available states associated to the coherent field.

In Fig. \ref{fig:Con_SB} we have finally shown the entanglement evolution for the two-qubit SB model.
Aiming at revealing general features of entanglement deterioration in this system, 
we have performed an ensemble average over $10^6$ samples applying Eq. (\ref{concuSBM})
with $x=1$, $\phi=\pi/4$ and random values of $g_j$ and $\omega_j$ taken from interval $[0,1]$.
As we see, increasing the environmental modes results in a faster decay of entanglement.
As expected, for short (long) times a Gaussian (exponential) behaviour is observed \cite{Maximilian}.
Due to both the initial Bell state of the central system and the dephasing local interaction,
there is no entanglement sudden death, as we can deduce from Eq. (\ref{concuSBM}).

\subsection{Concurrence-Purity diagram}
\begin{figure}[t!]
 \centering
 \includegraphics[width=8.5cm, height=5.5cm]{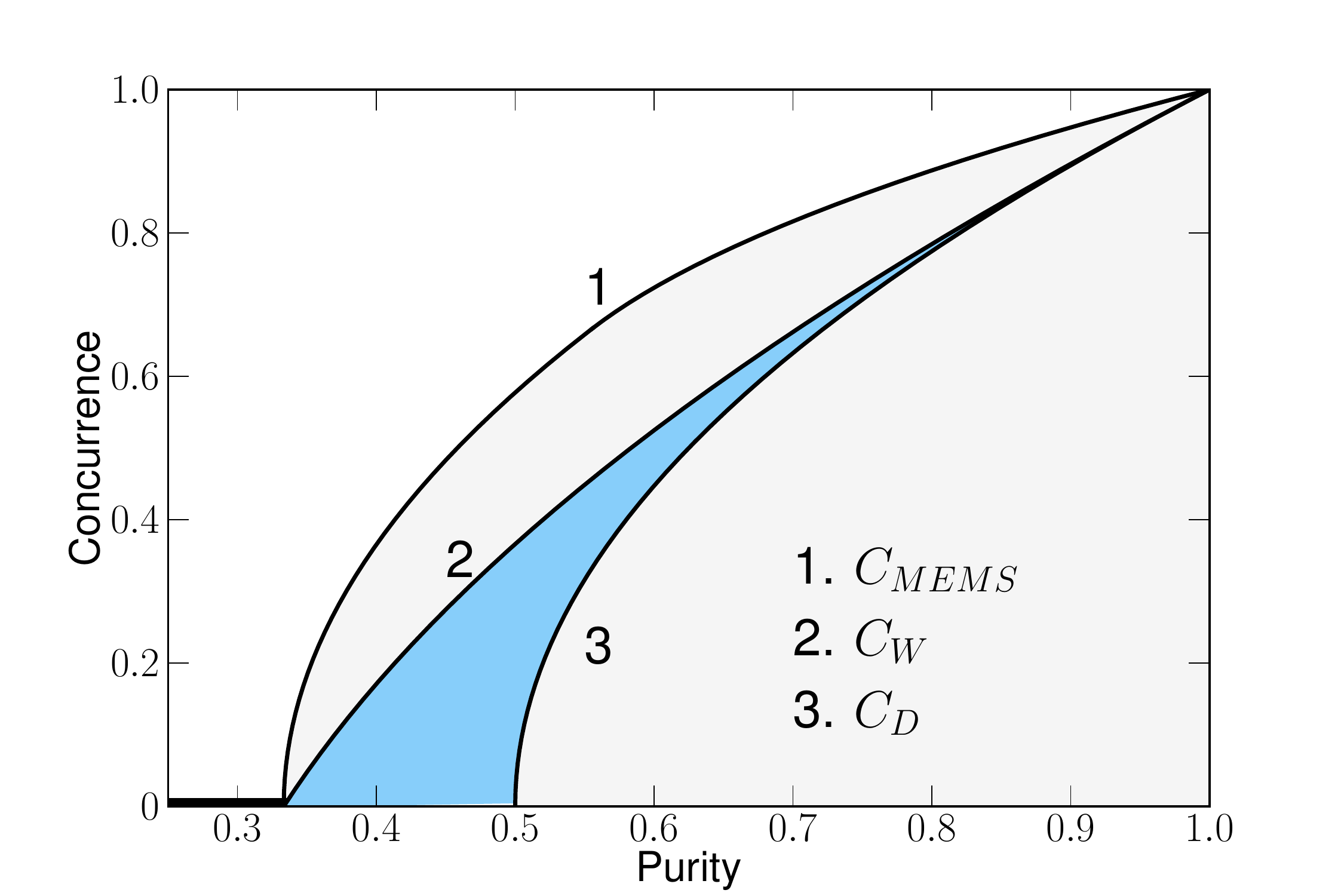}
 \caption{$C$-$P$ plane for two qubits. Curve 1 corresponds to maximally 
 entangled mixed states (MEMS). Curve 2 is for Werner states. The area
 coloured in blue is the region allowed for maximally entangled pure 
 two-qubit states when they are under the action of a unital quantum local
 channel. This region has a lower and upper bound given by $C_{D}$ and $C_W$ respectively \cite{Ziman-CP05}.}
 \label{cp:plane}
\end{figure}

A useful way to characterize bipartite quantum states is given by 
the concurrence-purity diagram or $C$-$P$ plane \cite{Ziman-CP05}.
In Fig.~\ref{cp:plane} we show for convenience a typical concurrence-purity diagram
specifying the relevant regions.
A point on this diagram gives the value of 
mixedness and entanglement at the same time. Those quantum states 
for which a definite value of  purity can reach the maximum degree of
entanglement are known as maximally entangled mixed states
(MEMS) \cite{Izaka00}. 
MEMS are represented by curve 1 ($C_{MEMS}$) in the $C$-$P$ plane. 
The area below the MEMS curve specifies the region of physical
quantum states.
Werner states ($\phi=\pi/4$ in Eq. (\ref{initial:state})) are depicted 
by curve 2 ($C_W$). Curve 3 ($C_D$) is given by Eq.~(\ref{C-P_SBM}) with $\phi=\pi/4$ which 
corresponds to  a decoherence process induced by a dephasing interaction.
\

In light of the dynamical results we have obtained for purity and concurrence,
we analyze their relation using the $C$-$P$ diagram. 
We first make some remarks about the nature of the quantum operations involved in our models.
We emphasize that the spectator configuration is a physical example
of a local quantum operation (channel) acting on a bipartite quantum
state (the state of the two-qubit central system). In this sense, 
environment performs operations (trough the interaction) on one of the two qubits. 
These local operations can be {\it unital} or {\it non-unital}. Unital channels are maps that
leave invariant the uniform state, i.e., the total mixture state. 
It is known \cite{Ziman-CP05} that initial Bell states under the action of  
unital channels lie in the region bounded by curves 2 and 3 in the $C$-$P$ plane (blue shadow) of Fig. \ref{cp:plane}.
Characterizing the behaviour of our quantum channels within this diagram is therefore desirable and can provide new overall insights on concurrence-purity dynamical relations. 
\begin{figure}[t!]
\centering
{\subfloat[]{\includegraphics[width=8.5cm, height=5.5cm]{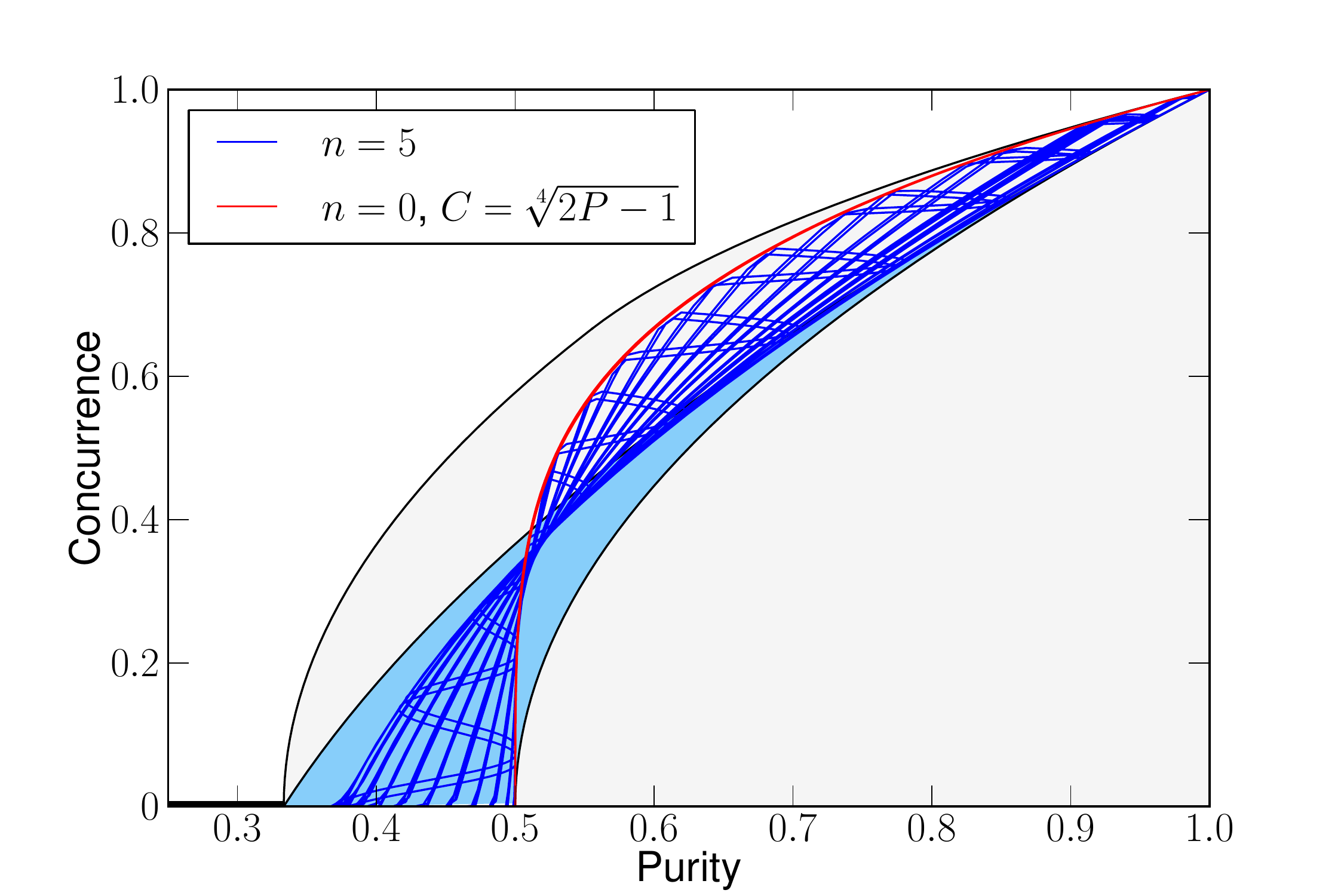}} \\
\subfloat[]{\includegraphics[width=8.5cm, height=5.5cm]{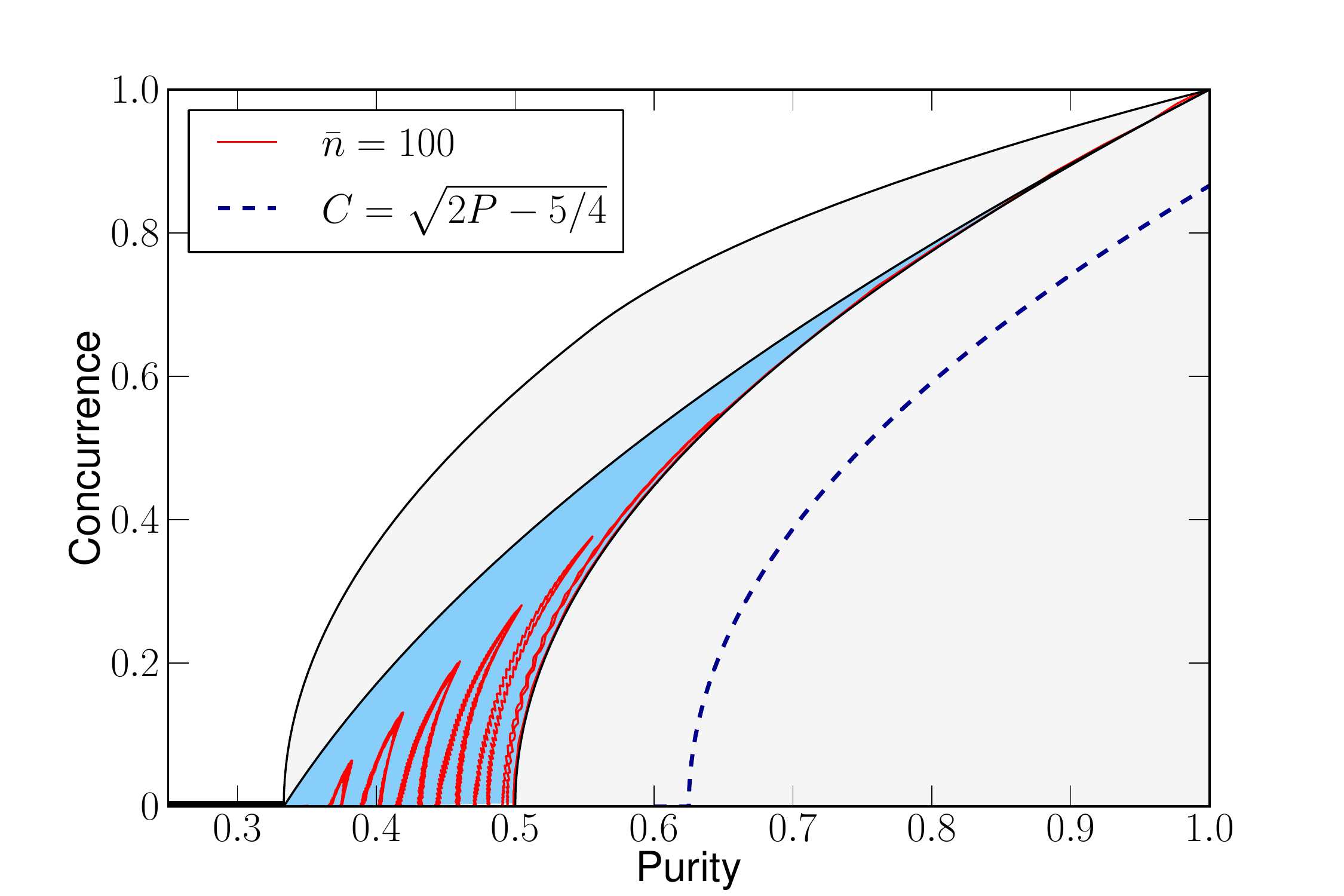}}}
\caption{$C$-$P$ plane representation for the local operation induced by Tavis-Cummings interaction.
 The central system starts in a Bell state and the field in: (a) Fock state $n$=$5$ (blue lines) and
 $n$=$0$ (red line), (b) coherent state $\bar{n}$=$100$ (red line). 
 Dashed line in (b) corresponds to a dephasing channel generated by the SB interaction (see text).}
\label{CP_plane_TC_5Fock}
\end{figure}

In Fig.~\ref{CP_plane_TC_5Fock}~(a) we show the behaviour of the channel acting on a Bell
state generated by the two-qubit TC dynamics in the spectator scheme. 
The starting point is the right upper corner in the plane.
Two representative cases for the initial state of the environment are shown: $i$) $n=0$ and $ii)$ $n=5$.
For the vacuum state a simple analytical relation between purity and concurrence can be obtained Eq.~(\ref{c-p::homogenization}):
$C=\sqrt[4]{2P-1}$ (red line), which for a long interval of time is outside of the unital region. 
This channel is related to the homogenization process describing exponential decay of correlations
in which the vacuum state is the fixed point of the dynamics. The case $n=5$ is shown in blue
and gives rise to a rich loop structure  due to immeasurability and non-Markovian behaviour in the evolution of
purity and concurrence. It must be mentioned that the associated  $C$-$P$ line for the vacuum state is also 
a loop over itself reaching zero entanglement at times $gt=(m+1/2)\pi$. 
These loops are exceptions to the rule that lines in the $C$-$P$ plane must be non-increasing 
if they are generated by Markovian semigroup dynamics. Hence, the appearance of this loops is 
due to the non-Markovian evolution considered in this work as we were able to obtain the exact
reduced density operator for the central system. It should be noted that similar results (not shown)
for the Buck-Sukumar interaction in the $C$-$P$ diagram can be obtained;
in contrast to the spectator two-qubit TC model, closed loops emerge due to the
$\pi$-periodicity in the purity and concurrence.   

At this point it is interesting to see the $C$-$P$ dynamics for an initial coherent state for the environment
using the results of subsection~\ref{sub::coherent_state}. For an average number of field excitation $\bar n=100$,
signatures of long-time entanglement revivals are obtained before their occurrence (see Fig.~\ref{CP_plane_TC_5Fock}(b)).
Almost all the action of the local operation is contained in the unital region except for a small part generated by 
the short time dynamics near to the upper right corner.
The corresponding $C$-$P$ representation for the spectator two-qubit SB dynamics 
is also shown in Fig.~\ref{CP_plane_TC_5Fock}(b) (dashed line) using the obtained
generalized expression in Eq.~(\ref{C-P_SBM}) with $\phi=\pi/6$. As expected we 
observe a typical decoherence process induced by dephasing, this process being represented as a rescaled $C_{D}$ curve.

\subsection{Operational use of the concurrence-purity relations}
We now briefly discuss on the possible usefulness to have quantitative relations between concurrence and purity for implementing some specific protocols. It is known that entanglement must overcome some quantitative thresholds, for a given value of state purity, in order to allow quantum processes, such as teleportation \cite{qinQIP,roarXiv}, entanglement swapping \cite{roaPRA} and entanglement percolation \cite{acin2007NatPhys}.  
Our results under specific dynamical conditions allow to only measure purity of the system state at a given time $t$ for obtaining the value of concurrence and then checking if it is sufficient for the desired task. Such a procedure will in turn provide the time regions within which the task can be performed. 

We focus on the recently reported concurrence threshold for entanglement necessary to realize a teleportation protocol with quantum speedup \cite{roarXiv}. Such a threshold is equal to $C_\mathrm{th}=(\sqrt{\rho_{22}}-\sqrt{\rho_{33}})^2$ in the case when the entangled state shared between the two parties is a X state, which is just the one we have during the evolutions here considered.
For instance, for the SB dephasing model, where $\rho_{22}(t)=\rho_{33}(t)$, one immediately gets $C_\mathrm{th}=0$ at any time. The system state can be thus exploited for teleportation until $C(t)>0=C_\mathrm{th}$, which in turn means whenever purity is above its minimum value $P(t)>P_\mathrm{th}\equiv\sin^4\phi+\cos^4\phi$ (see Eq.~(\ref{C-P_SBM})). For the plot of Fig.~\ref{CP_plane_TC_5Fock}(b) it must be $P(t)>5/8$. Instead, for the TC model with the vacuum field state and the two qubits initially prepared in a Bell state, the entanglement threshold is time-dependent, namely $C_\mathrm{th}(t)=(1/2)\sin^2(gt)$. Quantum teleportation is then achievable at those times such that $C(t)=|\cos(gt)|>C_\mathrm{th}(t)$, which in terms of state purity also means $P(t)>P_\mathrm{th}(t)\equiv [1+C^4_\mathrm{th}(t)]/2$ according to Eq.~\eqref{c-p::homogenization}.

\section{Conclusions}\label{Conclusions}
In this paper we have presented different exactly solvable models
for the dynamics of entanglement and purity of a simple two-qubit
central system. We have taken advantage of the spectator configuration, where a qubit is isolated, 
in order to realize a single local quantum operation
acting on a maximally entangled pure state. Furthermore, it allows for straightforwardly find the 
evolved two-qubit density matrix once the quantum map of the open qubit is known.
We have obtained explicit analytical expressions for purity, concurrence
and their dynamical relations (Eqs.~\ref{c-p::homogenization} and \ref{C-P_SBM}) using  
Tavis-Cummings, Buck-Sukumar and spin-boson type interactions.
Our results confirm that even in the spectator scheme the entanglement can disappear at a finite time depending on the initial conditions, as previously found in other open quantum systems \cite{Cui07,lofrancoQIP,LoFrancoNatCom}.
Long-time entanglement revivals appear when a coherent state of
the radiation environment is considered, showing that even simpler systems that the ones 
treated in previous works~\cite{aolitareview,lofrancoreview,Yonac1,Yonac2} can reveal general features of 
entanglement evolution. 
In fact, the qualitative behaviors of the dynamics of quantum correlations, like entanglement, are analogous for bipartite systems of both open qubits and only one open qubit provided that the qubits are independent and locally interacting with their own environment.   

As a further source of information we have exploited
the $C$-$P$ diagram to characterize how local actions ruled by the environment 
affect an initial two-qubit Bell state. For the TC and BS interactions, the two-qubit state can reach
points outside of the unital region which thus evidences the non-unital nature of
these kind of quantum maps commonly employed in the context of quantum optics.
We have also discussed the potentiality of having concurrence-purity dynamical relations to assess quantitative entanglement and purity thresholds at a given time which allow specific quantum tasks, such as teleportation.   

These results motivate further studies of dynamical characterization of thresholds of purity and entanglement for implementing processes like entanglement swapping \cite{roaPRA} and entanglement percolation \cite{acin2007NatPhys}. 
For future works, it would be also interesting to consider more realistic models 
in the spirit of the spectator configuration, for instance introducing spontaneous
emission and cavity photon losses by means of Lindblad master equations.

\section*{Acknowledgements}
CGG, RRA and DE would like to express their gratitude to CONACyT for
financial support under scholarships No. 385108, 379732 and 413926, respectively.

\newpage
\section*{References}


\providecommand{\newblock}{}

\end{document}